# Compact Graph Model of Handwritten Images: Integration into Authentification and Recognition

Denis V. Popel

Department of Computer Science,
Baker University, Baldwin City,
KS 66006-0065, USA,
`popel@ieee.org`

**Abstract.** A novel algorithm for creating a mathematical model of curved shapes is introduced. The core of the algorithm is based on building a graph representation of the contoured image, which occupies less storage space than produced by raster compression techniques. Different advanced applications of the mathematical model are discussed: recognition of handwritten characters and verification of handwritten text and signatures for authentification purposes. Reducing the storage requirements due to the efficient mathematical model results in faster retrieval and processing times. The experimental outcomes in compression of contoured images and recognition of handwritten numerals are given.

## 1 Introduction

It is essential to reduce the size of an image, for example, for analysis of features in automatic character recognition or for creation of a database of authentic representations in secure applications. One can estimate that the amount of memory needed to store raster images is greater than for vector ones [5]. The problem arises if the contoured image like handwritten text or signature is scanned and only bitmap of this shape is produced. Hence, many algorithms were developed in the past years to compress raster images (LZW, JPEG,..., etc.), as well as for vector representation [3],[8]. From the other side, the extensive growth of portable devices requires developing techniques for efficient coding and transmitting audio-video information through wireless networks. Therefore, the simplification of graphical data is the first essential step towards increasing data transmission rate and enhancing services of cell phones and PDA computers [2]. The structural model discussed in this paper allows compressing images and creates a range of features to start automatic processing of contoured images, e.g. recognition.

Authentification and personal identification of financial documents are main tasks that guarantee sufficient safety in business activities. Nowadays there is a growth of crime in the area of forgery of signatures and falsifying handwritten documents. The current approach of solving this problem is using semiautomatic/manual client identification systems based on signature and handwritten text identification on personal cheques, credit card receipts, and other documents. Often only signatures and handwritten text can transform an ordinary paper into legal document. It would be ideal if the client personal characteristics like signature and handwriting could be identified by computer in full automatic mode [4],[7]. Now manual visual comparison of customer signatures and handwriting is widely used, where images are represented from the customer's document and from a database. For automatic authentification and for effective storage in a computer database, the new methods should be introduced which will lead to proper utilization of human and computing resources. This paper presents such an algorithm to build a mathematical model of handwritten characters and signatures and shows how to compress any contoured images. The reduced amount of memory is required to store this model which can be generally used for automatic analysis and identification of images, for instance, during automatic identification of a person using signature or recognition of handwritten text [1],[7].

## 2  Identification Process as a Motivation of Our Study

In a process of checking of signature or handwritten text authenticity from the paper document, a system should be equipped with the following components: (i) *a scanner* with image retrieving program; (ii) *a program* to compress images, which is integrated with a database management system; (iii) *a program* to reproduce compressed image.

The process of addition of an authentic signature or handwriting to the database consists of the following steps: scanning of image from a document, building mathematical model, and placing it as a separate record in the database. Afterwards the model is extracted by some database key (for example, by client account number in bank applications), and the reproduction program restores the original view of the signature or handwriting. The image to be examined should be also scanned and displayed on the screen. Thus, a bank clerk has an opportunity to access the original client signature or handwriting and to compare it with the current image. The information stored in such way is insufficient to match authentification expertise, which required from 6 to 20 original objects (depending on signature's or handwriting complexity). In this paper, we propose an algorithm to construct generalized mathematical model that stores essential and invariable features for a person and can be used as a basis to provide automatic authenticity confirmation of handwritten text or signatures.

## 3  Graph Model

In our approach, the transformation of raster image consists of the following stages: (1) image thinning; (2) image representation as a graph; (3) shape smoothing; (4) graph minimization; (5) shape compression.

### 3.1 Image Thinning

Skeletonization is an iterative procedure of bitmap processing, when all contour lines are transformed to singular pixel ones. The modified Naccache - Shinghal algorithm [6] is used in the described approach. The modification utilizes arithmetic differential operators to find the skeleton and provides good results for handwritten text and signatures. The average width of lines is measured during the skeletonization stage.

### 3.2 Transformation of Bitmap to Graph-like Representation

We introduce the notation *pixel index* which is the number of neighboring image pixels in $3 \times 3$ window around the current point. Pixel with index 1 named endpoint (the beginning or the end of curve), with index 2 - line point, and pixels with indexes 3 and 4 are nodes points (junction and crossing of lines). The graph is represented in memory using two lists: the list of nodes and the list of the descriptors of contour branches. The element of the list of nodes contains $X,Y$ coordinates of a bitmap pixel and pointers to contour branches, which started from this node. The descriptor of contour branch is a *chain code* (Freeman code [3]), where each element carries 8-bit code of the next contour direction (see Figure 3(a) for details). At this stage, the thinned bitmap is transformed to a graph description, where nodes are pixels with indexes 1,3 or 4, and arcs are contour branches. The starting node for the looped contours is selected arbitrary.

### 3.3 Shape Smoothing

As the scanned image has different distortions and noise, thinning shape contains some false nodes. These false nodes (treated as defects) do not have serious influence on the quality of the restored image, but their description requires a lot of additional memory. The shape is smoothed to eliminate these defects. This operation (a,b) erases false nodes, (c) smoothes of all broken contours, (d) eliminates all nodes with index 4, and (e) erases some nodes with index 3 (Figure 1). The stage of smoothing can be omitted if the lossless representation is required by applications. The suggested algorithm supports both lossy and lossless representation strategies. Eliminating nodes with index 4 does not change the shape of processing image (lossless strategy), and after the smoothing stage, graph contains only nodes with index 1 and 3.

### 3.4 Graph Minimization

The graph description obtained at the second stage is redundant, therefore it is possible to conduct graph minimization by eliminating some nodes and connecting corresponding contour branches. Finally, each branch has only one $j$-th node with index $I_j$ assigned to the beginning of the chain of pixels.

**Lemma 1.** *Before minimization, each branch of the graph interconnects two nodes with indexes 1, 3 or 4.*

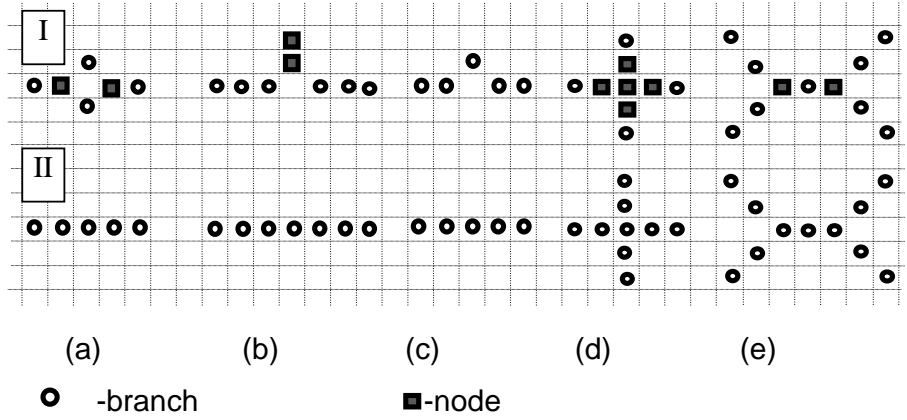

**Fig. 1.** Smoothing of image: (I) source shape, (II) smoothed shape

**Theorem 1.** *The number of branches B of the graph with no loops can be determined as follows:*
$$B = \frac{I_1 + I_2 + ... + I_N}{2}, \quad (1)$$
*where N is the number of nodes of the source graph, $I_j$ - the index of j-th node (excluding nodes with index 4).*

**Corollary 1.** *It follows from the Theorem 1 that the number of branches for the looped contour is*
$$B = \frac{N_1 + 3 \cdot N_3}{2} + N_{loops}, \quad (2)$$
*where $N_1$ is the number of nodes with index 1, $N_3$ - number of nodes with index 3, and $N_{loops}$ is the number of arbitrary selected nodes with index 1 to cover all loops.*

Graph minimization can eliminate nodes with indexes 1 and 3. It minimizes the number of branches in the graph while reducing number of nodes with index 3. The Corollary 1 can be reformulated for the minimal number of nodes and branches.

**Corollary 2.** *The number of branches in minimized graph equals*
$$B^{min} = \frac{N_1^{min} + 3 \cdot N_3^{min}}{2} + N_{loops}, \quad (3)$$
*where $N_1^{min}$ and $N_3^{min}$ are the minimal number of nodes with indexes 1 and 3 correspondingly.*

The graph minimization technique resolves spanning tree problem through the following steps:

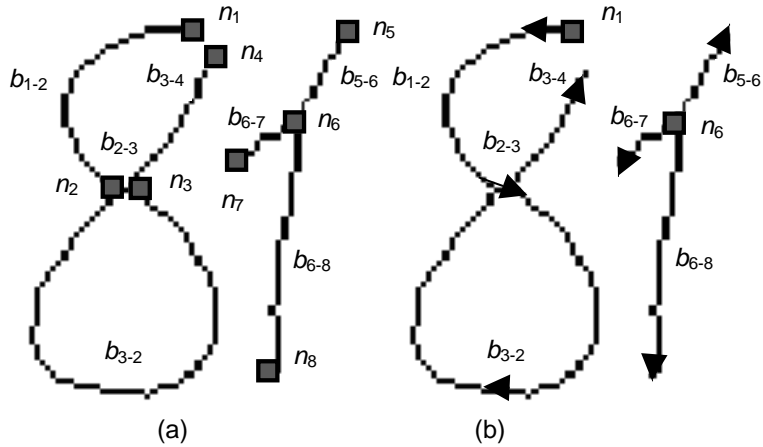

**Fig. 2.** Handwritten digits as a graph: (a) before smoothing and minimization; (b) after minimization (lossless smoothing)

**Step 1.** `Select a node with index 3. Trace three uncovered outgoing`
   `branches following links from the selected node until tracing`
   `reaches already covered node or a node with index 1. Mark the`
   `node and all branches as covered.`
**Step 2.** `Repeat Step 1 until all nodes with index 3 are covered.`
**Step 3.** `Select a node with index 1 and trace outgoing branch. Mark`
   `the node and branch as covered.`
**Step 4.** `Repeat Step 3 until all nodes with index 1 are covered.`

*Example 1.* Figure 2 shows the connected graph with eight nodes $n_1, \ldots, n_8$ and seven branches $b_{1-2}, b_{2-3}, \ldots, b_{6-8}$. Nodes $n_2$ and $n_3$ are removed by smoothing operation, and the branches $b_{1-2}, b_{2-3}, b_{3-2}$ and $b_{3-4}$ are transformed into one branch which covers the entire contour. Before minimization the graph description contains six nodes and four branches. Minimal graph has two nodes $n_1, n_6$ and four branches. So the number of nodes is reduced from eight to two.

### 3.5 Contour Compression

After thinning and smoothing stages, the contour does not have sharp shifts and distortions. Therefore the next contour point in the branch description has only three positions related to the current one (see Figure 3(b)). This property allows us to represent the branch dynamically using relative coordinates $(-1, 0, 1)$, which provide $L \cdot \log_2 3$ ($1.6 \cdot L$) bits instead of $8 \cdot L$ bits for the chain of length $L$. Thus, the size of the branch description can be reduced. In addition, all branches are compressed by modification of widely used *Run-Length Encoding algorithm*.

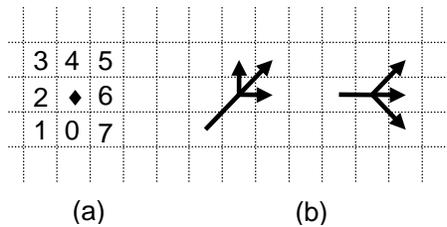

**Fig. 3.** Contour directions

## 4 The Reproduction of Image

The process of reproduction is based on mathematical vector model and should be executed in two stages: skeleton reproduction; contour thickening. The skeleton reproduction is fulfilled by traversing through graph nodes and transforming of contour branches into bitmaps. Obviously the described method does not allow us to reproduce the initial image with pixel accuracy (Figure 4). However as analysis shows this aspect does not influence on image comparison in the context of the authenticity problem as well as character recognition task discussed above. If exact image reproduction is necessary, it is possible to use a modified description, where line width is assigned to each point of the contour.

## 5 Experimental Results

The algorithm described above is realized and experiments were conducted on signatures, handwritten text (various languages) and contoured images.

In the first series of experiments, the described above algorithm was used directly to resolve the storage problem. Experiments show that the compression ratio for (a) signatures is 10-20, (b) handwritten text – between 5 and 12, and (c) curved shapes – between 8 and 17, (see Table 1). In some cases, the results exceed a compressing degree produced by well-known archiving programs. For example, the initial image with handwriting in BMP format has 1962 bytes, compressed by ZIP - 701 bytes, compressed using presented algorithm - 154 bytes. The contour compression ratio is 11.9, that is 4.3 times effective than using raster compression. Figure 5 compares the result of our algorithm with the outcome of CorelTrace program.

Second series of experiments covers recognition of handwritten numerals. Twenty distinctive features where extracted from the mathematical model: relative distances, number of different nodes, number of loops, straight strokes,..., etc. These features are invariant to rotation between −45 and 45 degrees and scaling. The experiments were fulfilled on MNIST[1] database. The estimated error rate of recognition algorithm is about 5.9%.

---
[1] http://www.research.att.com/~ yann/exdb/mnist/index.html

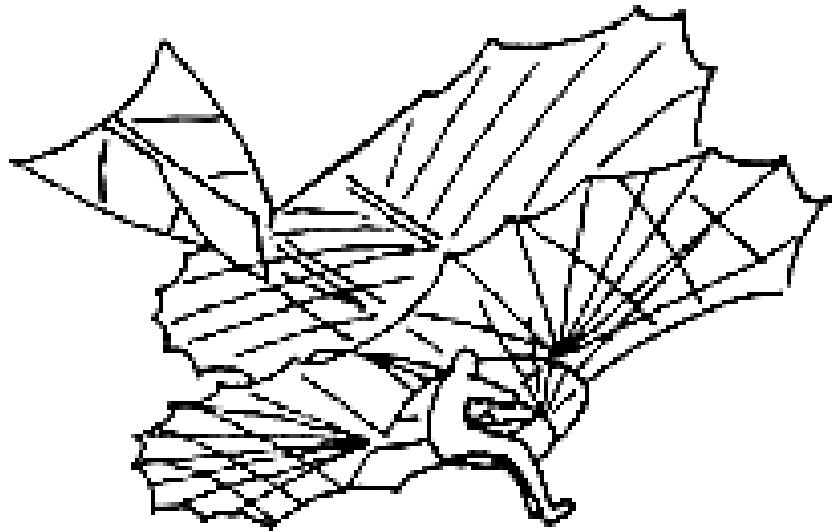

(a)

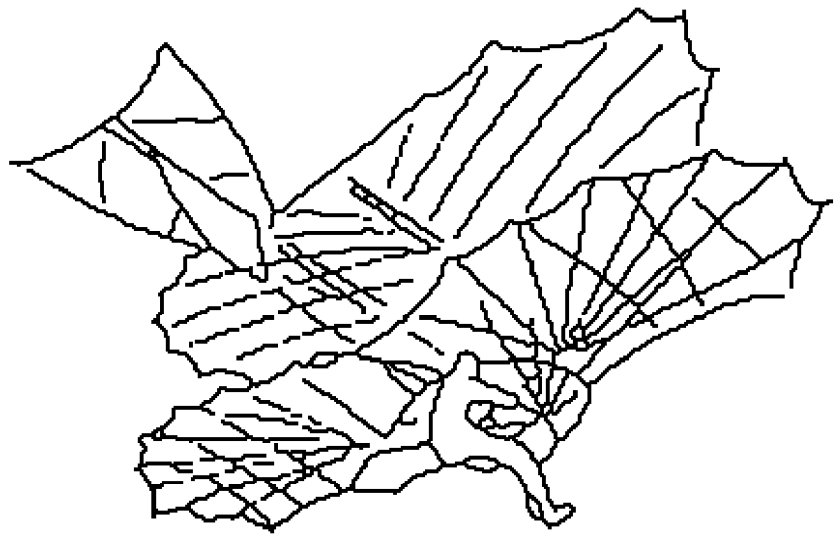

(b)

**Fig. 4.** Example of lossy compression and reproduction: (a) initial image – Da Vinci drawing, (b) reconstructed image

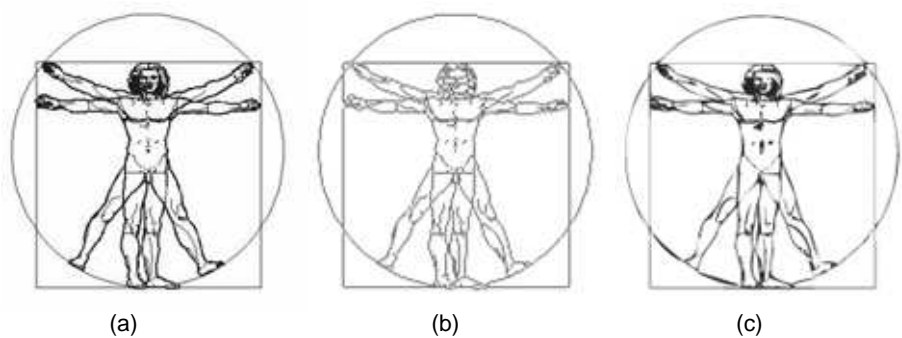

**Fig. 5.** Da Vinci drawing: (a) its original shape, (b) processed by suggested algorithm and (c) vector representation obtained by CorelTrace program

## 6 Concluding Remarks and Ongoing Research

In this paper, we presented a novel algorithm to create a generalized mathematical model for contoured images. Several related issues, such as automatic authenticity confirmation and recognition of handwritten characters based on this model, were discussed. We are planning to extend the proposed mathematical model of contoured images and integrate it in authentification systems. This extension will reflect probability characteristics of image attributes and structural features of handwritten objects. Another feasible application of the graph model is the compression of large *Computer Aided Design* (CAD) and *Geographic Information System* (GIS) images.

## References

1. Al-Emami, S., Usher, M.: On-Line Recognition of Handwritten Arabic Characters, IEEE Trans. on Pattern Analysis and Machine Intelligence, vol. 12, no. 7 (1990) 704–709
2. Ansorge, M., Pellandini, F., Tanner, S., Bracamonte, J., Stadelmann, P., Nagel, J.-L., Seitz, P., Blanc, N., Piguet, C.: Very Low Power Image Acquisition and Processing for Mobile Communication Devices, Proc. IEEE Int. Symp. on Signals, Circuits and Systems - SCS'2001 (2001) 289–296
3. Freeman, H.: On the Encoding of Arbitrary Geometric Configurations. IEEE Trans. Elect. Computers, vol. ES-10 (1961) 260–268
4. Gazzolo, G., Bruzzone, L.: Real Time Signature Recognition: A Method for Personal Identification, Proc. Int. Conf. on Document Analysis and Recognition (1993) 707–709
5. Gilewski, J., Phillips, P., Popel, D., Yanushkevich, S.: Educational Aspects: Handwriting Recognition - Neural Networks - Fuzzy Logic, Proc. IAPR Int. Conf. on Pattern Recognition and Information Processing - PRIP'97, vol. 1 (1997) 39–47
6. Naccache, N.J., Shingal, R.: SPTA: A Proposed Algorithm for Thinning Binary Patterns, IEEE Trans. Systems. Man. Cybern., SMC - 14, no. 3 (1984) 409–418

Table 1. Compression characteristics (images were scanned at 100 dpi)

| Initial image and size in bytes | Restored image and compressed size in bytes | Compression rate |
|---|---|---|
| 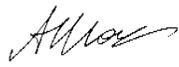 3742 | 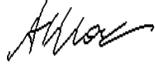 193 | 19.3 |
| 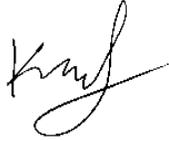 4158 | 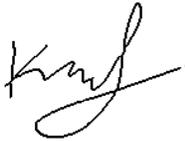 210 | 19.8 |
| 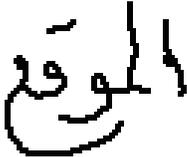 914 | 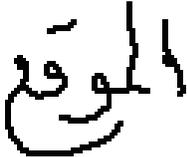 157 | 5.82 |
| 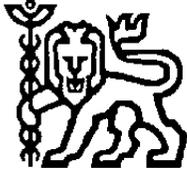 10021 | 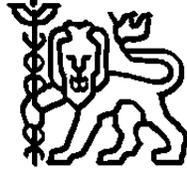 683 | 14.67 |

7. Popel, D., Ali Muhammed, T., Hakeem, N., Cheushev, V.: Compression of Handwritten Arabic Characters Using Mathematical Vector Model, Proc. Int. Workshop on Software for Arabic Language as a part of IEEE Int. Conf. on Computer Systems and Applications (2001) 30–33
8. Song, J., Su, F., Chen, J., Tai, C. L., Cai, S.: Line net global vectorization: an algorithm and its performance analysis, Proc. IEEE Conf. on Computer Vision and Pattern Recognition (2000) 383–388